\documentclass[a4paper,11pt]{article}
\usepackage{jcappub}
\usepackage{amsmath}
\usepackage{graphicx}
\usepackage{enumitem}
\usepackage{hyperref}

\makeatletter
\gdef\@fpheader{}
\makeatother

\newlength{\fullw}
\newlength{\halfw}
\newlength{\twofigw}
\newlength{\onefigw}
\newcommand{\Mpc}{\mathrm{Mpc}}

\newcommand{\order}[1]{\mathcal{O}\!\left(#1\right)}
\newcommand{\boldmathsymbol}[1]{{\ensuremath{\boldsymbol{#1}}}}
\newcommand{\bm}[1]{\boldmathsymbol{#1}}
\newcommand{\cross}[2]{(\textrm{#1$\times$#2})}

\newcommand{\SONG}{\textsc{Song}}
\newcommand{\CLASS}{\textsc{Class}}
\newcommand{\CAMB}{\textsc{Camb}}
\newcommand{\LCDM}{$\Lambda\textrm{CDM}$}

\newcommand{\calC}{\mathcal{C}}

\newcommand{\dd}{{\mathrm{d}}}
\newcommand{\uini}{\mathrm{ini}}
\newcommand{\ub}{\mathrm{b}}
\newcommand{\ue}{\mathrm{e}}

\newcommand{\xe}{x_\ue}
\newcommand{\xet}{\tilde{\xe}}
\newcommand{\uSZ}{\mathrm{SZ}}
\newcommand{\uEoR}{\mathrm{EoR}}
\newcommand{\uGR}{\mathrm{GR}}
\newcommand{\pure}{\mathrm{pure}}
\newcommand{\loss}{\mathrm{loss}}
\newcommand{\rec}{\mathrm{rec}}
\newcommand{\blur}{\mathrm{blur}}
\newcommand{\lens}{\mathrm{lens}}

\newcommand{\etaini}{\eta_{\uini}}

\title{CMB anisotropies from patchy reionisation and diffuse
  Sunyaev-Zel'dovich effects}

\author[]{Christian Fidler}
\author[]{and Christophe Ringeval}

\affiliation[]{Centre for Cosmology, Particle Physics and Phenomenology,
  Institute of Mathematics and Physics, Louvain University, 2 Chemin
  du Cyclotron, 1348 Louvain-la-Neuve, Belgium}
  
\emailAdd{christophe.ringeval@uclouvain.be}
\emailAdd{christian.fidler@uclouvain.be}

\date{today}

\begin{document}

\abstract{Anisotropies in the Cosmic Microwave Background (CMB) can be
  induced during the later stages of cosmic evolution, and in
  particular during and after the Epoch of
  Reionisation. Inhomogeneities in the ionised fraction, but also in
  the baryon density, in the velocity fields and in the gravitational
  potentials are expected to generate correlated CMB perturbations. We
  present a complete relativistic treatment of all these effects, up
  to second order in perturbation theory, that we solve using the
  numerical Boltzmann code {\SONG}. The physical origin and relevance
  of all second order terms are carefully discussed. In addition to
  collisional and gravitational contributions, we identify the diffuse
  analogue of the blurring and kinetic Sunyaev-Zel'dovich (SZ)
  effects.  Our approach naturally includes the correlations between
  the imprint from patchy reionisation and the diffuse SZ effects
  thereby allowing us to derive reliable estimates of the induced
  temperature and polarisation CMB angular power spectra. In
  particular, we show that the $B$-modes generated at intermediate
  length-scales ($\ell \simeq 100$) have the same amplitude as the
  $B$-modes coming from primordial gravitational waves with a
  tensor-to-scalar ratio $r=10^{-4}$.}


\maketitle

\section{Introduction}

\label{sec:intro}

The cosmic microwave background (CMB) has been exploited with great
success in the past decades and remains the main pillar of precision
cosmology. Beyond the linear physics in the era of recombination, a
large range of secondary effects are present which are crucial for
the analysis of current and future missions~\cite{Matsumura:2013aja,
  Adam:2016hgk, 2016arXiv161208270C, Aghanim:2017bzn,
  2017arXiv170404501R, 2017arXiv170604516D}. In addition to constituting a
``background'' to the purely linear physics, they also carry information
about the cosmic evolution itself~\cite{Deutsch:2017ybc}. Secondary
effects appear both at linear and non-linear order in the theory of
the cosmological perturbations. The linear ones are the late
Integrated Sachs-Wolfe effect (ISW) and Doppler effect from
collisions after reionisation. These ones are implemented in
most linear Boltzmann codes currently available and will not be
discussed in the following~\cite{Lewis:1999bs, 2011arXiv1104.2932L,
  Blas:2011rf}. Non-linear secondaries can be separated into
different classes:
\begin{enumerate}[label={\arabic*)}]
\item Gravitational dynamics. General Relativity induces CMB
  perturbations going beyond the linear ISW and SW effects. These are
  called the Rees-Sciama effects~\cite{rees1968large} and are
  typically negligible compared to the remaining
  secondaries~\cite{Seljak:1995eu, Cooray:2002ee}.
	
\item Line-of-sight distortions. Gravitational effects bend the light
  rays. This includes the gravitational
  lensing~\cite{blanchard1987gravitational, kashlinsky1988small,
    Lewis:2006fu}, but also time-delay~\cite{Seljak:1994wa} and
  redshift effects\footnote{These describe the equivalent of the ISW
    effect but now acting on the linear photon perturbations instead
    of the background.} as discussed in Refs.~\cite{Huang:2012ub,
    Pettinari:2013he, Fidler:2014zwa}. Lensing is an important and
  observable effect, while the latter ones are
  subleading~\cite{saito2014geodesic, Fidler:2014zwa}.
	
\item Collisional dynamics at recombination. Contributions
  beyond the linear collision term, as for example higher order
  phase-space
  enhancements~\cite{Su:2012gt, Huang:2012ub, Pettinari:2013he}. The
  effect is of intermediate size and is slightly below the Planck
  satellite detection sensitivity~\cite{Pettinari:2014iha}.
	
\item Collisional dynamics at the Epoch of Reionisation. The same
  physics at work during recombination becomes relevant again during
  the Epoch of Reionisation (EoR)~\cite{Loeb:2000fc, Barkana:2006ep}. Since
  reionisation is a completely inhomogeneous process,
  the impact of second-order corrections is enhanced. The two main
  contributions are the blurring of existing CMB anisotropies due to
  the ionised gas~\cite{Dvorkin:2008tf} and the induction of
  perturbations by collisions~\cite{Hu:1999vq, Dore:2007bz}. This
  signal is expected to be a direct probe of the EoR~\cite{Smith:2016lnt}.
	
\item Collisional dynamics after EoR. Once reionisation is completed,
  the Universe is fully ionised and CMB photons may scatter with the
  free electrons. Collisions are most likely in overdense regions
  and the resulting imprints in the CMB are known as the
  Sunyaev-Zel'dovich (SZ) effects. As for the reionisation
  contribution, we distinguish between the less studied blurring
  effects~\cite{HernandezMonteagudo:2009ma} and the well-known
  collision-induced CMB perturbations~\cite{Sunyaev:1980,
    Ostriker:1986, Jaffe:1997ye}. Of especial interest is the
  generation of polarisation from the SZ
  effects~\cite{Sazonov:1999zp}. Let us stress that, in our case,
  collisions are not only occurring in dense galactic clusters but are
  diffuse over the whole Universe.
\end{enumerate}

In this paper we attempt to include these CMB secondaries in the
theory of the cosmological perturbations at second order. Such a
treatment has originally been proposed by Hu, Scott and Silk in
Ref.~\cite{Hu:1993tc}, but this work was dedicated to the small scale
temperature anisotropies, considering only the most dominant
contributions. Our work therefore extends these results to the full
sky, encompassing the entire range of second order sources and, more importantly,
to the $E$- and $B$-mode polarisation.
  
In the past years several second-order Boltzmann codes have been
developed~\cite{Su:2012gt, Huang:2012ub, Pettinari:2013he} that can
accurately compute the effects discussed in point 1) and 3), plus the
redshift effects from point 2). Lensing is already treated in linear
codes such as {\CAMB}~\cite{Lewis:1999bs} or
{\CLASS}~\cite{Blas:2011rf}. However, the terms discussed in the
points 4) and 5) have not be considered. We focus on their
implementation in the Boltzmann code {\SONG}~\cite{Pettinari:2013he},
in full General Relativity and only employing the assumption of
neglecting third and higher order corrections.

The paper is organised as follows. Key results from second-order
perturbation theory are reviewed in section~\ref{sec:2pert} while we
detail our numerical implementation in
section~\ref{sec:num}. Numerical calculations of the temperature and
polarisation power spectra are presented in
section~\ref{sec:res}. We conclude in section~\ref{sec:con}.

\section{Second order perturbation theory}
\label{sec:2pert}

In the following we introduce our notation and review some results
from the theory of the cosmological perturbations at second-order.

\subsection{Notation}

Greek indices $\mu, \nu$ label a relativistic 4-vector, while indices
$\alpha,\beta$ describe a special relativistic 4-vector (for example
in the local inertial frame). Latin indices $i,j$ refer to the
spatial parts of both relativistic and special relativistic vectors,
while the temporal index is labelled by a $0$. Latin indices $a,b$ are
helicity indices and run over $a = \pm$ for photons.
   
The metric is assumed to be of the form
\begin{equation}
ds^2 = a^2\left\{(1+2A)\dd \eta^2 + 2 B_i \dd\eta \dd x^i -
\left[(1+2D)\delta_{ij} + 2 E_{ij}\right]\dd x^i \dd x^j\right\},
\end{equation}
with the lapse perturbation $A$, the shift $\bm{B}\equiv\{B^{i}\}$,
the spatial trace perturbation $D$ and the symmetric spatial tensor
perturbation $E_{ij}$. Einstein equations, stress tensor and Boltzmann
equations are then expanded to second-order in perturbation theory $X
= X^{(1)} + X^{(2)} + \cdots$. We further assume that $B_i^{(1)}=
E^{(1)}_{ij} = 0$, corresponding to the so-called Poisson gauge at
linear order, plus the assumption that vector and tensor modes, for
example due to primordial gravitational waves, are counted as a
second-order perturbation.

The stress tensor is decomposed as
\begin{equation}
T_{\mu\nu} = (\rho+P)u_\mu u_\nu - P g_{\mu\nu} + \Sigma_{\mu\nu} \,,
\end{equation}
where $\rho$ is the density, $P$ the pressure, $\Sigma_{\mu\nu}$ the
anisotropic stress tensor and $u_\mu$ the rest-frame 4-velocity.
 
We characterise the species in the Universe by their distribution
function $f_{ab}(\eta,\bm{x},\bm{p})$; the probability of finding a
particle of a given species with 3-momentum $\bm{p}$ at the position
$\bm{x}$ and conformal time $\eta$.  The stress tensor can be built
from the distribution function by evaluating the first kinetic moments
\begin{equation}
T_{\mu\nu} = [e^\alpha]_\mu [e^\beta]_\nu \int \dfrac{\dd \bm{p}}{(2\pi)^3}f(\eta,\bm{x},\bm{p}) \frac{p_\alpha p_\beta}{p_0}\,,
\end{equation}
where $[e^\alpha]_\mu$ stands for the tetrad and $p_0$ is the
rest frame energy.

Cold species, such as massive particles, have a trivial phase-space
and all information is contained in the first moments: the density
$\rho$ and the 3-velocity $\bm{v}$. The latter being defined as the
spatial part of $u$ at linear order $v^{(1)}_i \equiv a u^{(1)}_i$. Relativistic species, however, require the knowledge of
the distribution function. When not interested in the spectral
information, one may integrate over the momentum and we define
\begin{equation}
\Delta_{ab}(\eta,\bm{x},\bm{n}) \equiv \dfrac{\int \dd q q^3
  f_{ab}(\eta, x, q \bm{n})}{\int \dd q q^3 f_{I}^{(0)}(q)}\,,
\end{equation}
where $q_i= a p_i$ is the comoving momentum of direction
$\bm{n}$ and magnitude $q$. These quantities are sufficient to
evaluate the stress tensor and to provide a closed set of
equations. The $\Delta_{ab}$ can be related to the temperature of
an equivalent blackbody spectrum.

We use the Fourier conventions
\begin{equation}
  A(\bm{x}) = \int \frac{\dd^3 \bm{k}} {(2\pi)^3} e^{i\bm{k} \cdot \bm{x}}
  A(\bm{k})\,,
\end{equation}
while all multiplications involving different wavenumbers in Fourier
space are implicitly assumed to be convolution integrals, namely
\begin{equation}
A(\bm{k}_1) B(\bm{k}_2) \equiv  \int \dfrac{\dd^3\bm{k}_1}{(2\pi)^3}\int \dfrac{\dd^3\bm{k}_2}{(2\pi)^3}(2\pi)^3\delta^3(\bm{k}-\bm{k}_1 -\bm{k}_2)A(\bm{k}_1)B(\bm{k}_2)\,.
\end{equation}
Multipole decomposition is assumed to be a decomposition over the
spin-weighted spherical harmonics as
\begin{equation}
f_{ab}(\eta,\bm{x},\bm{q}) = \sum \limits_{l,m} (-i)^l
\sqrt{\frac{4\pi}{2l+1}}f_{ab,lm}(\eta,\bm{k},q)Y_{lm}^s(\bm{n}) \,.
\end{equation}
Here, $s$ is the spin associated with the helicity state $ab$ and
$Y_{lm}^s$ are the spin-weighted spherical harmonics. Note that $ab$
is usually decomposed into the Stokes parameters $X=I,V,Q,U$ before
multipole decomposition, or into $X = I,V,E,B$ after multipole
decomposition. 

Similarly to our notation for the convolutions, an implicit summation at fixed $m$ is
assumed over all multipole coefficients labelled with $m_1$ or
$m_2\equiv m-m_1$. Finally, indices within square bracket refer to
helicity states. Further details are given in
Refs.~\cite{Beneke:2010eg, Beneke:2011kc}.

\subsection{Line-of-Sight integration}

The line-of-sight integration is an important tool for solving the
Boltzmann equations. It analytically describes the generation of the
large present-day multipoles from sources active at much lower
multipoles. Thereby it makes possible the truncation of the Boltzmann
hierarchy at a given multipole $l$; higher multipole moments being
compensated by using an analytical
expression~\cite{Seljak:1996is,Hu:1997hp}. Let us assume a
differential equation in Fourier space of the form
\begin{equation}
\dot{\Delta}_{ab} + i \bm{n} \cdot \bm{k} \Delta_{ab} = -|\dot{\kappa}|\Delta_{ab} 
+\rho_{ab}\,,
\end{equation}
where the term $ i \bm{n} \cdot \bm{k} \Delta_{ab}$ describes the free
propagation of photons, $|\dot{\kappa}|$ is the chance for
interactions and consequently $ -|\dot{\kappa}|\Delta_{ab}$ describes
collisions knocking the photons out of our line-of-sight. The
rightmost quantity, $\rho_{ab}$, contains all remaining terms that are found in
the full second-order equations. The solution to this equation is
formally given by
\begin{equation}
  \label{eq:los}
\Delta_{ab}(\eta,\bm{k},\bm{n}) = \int \limits_{\etaini}^{\eta}
\dd \eta' e^{- i \bm{n} \cdot \bm{k} (\eta - \eta') - \kappa(\eta,\eta')}
\rho_{ab}(\eta',\bm{k},\bm{n})\,,
\end{equation}
where $\kappa$ is the integrated optical depth and the exponential
$e^{- i \bm{n} \cdot \bm{k} (\eta - \eta')}$ describes the generation
of a more complex angular distribution by the free propagation of
photons. The multipole decomposition of this equation is non-trivial
and can be found in Ref.~\cite{Beneke:2010eg}.

\subsection{Collision terms at second order}

The polarised second-order Boltzmann equations have been derived in
Refs.~\cite{Pitrou:2008hy, Beneke:2010eg} and contain numerous
sources, including the lensing and gravitational ones. In the
following, we explicitly write the source terms that are only due to
interactions, namely, the ones appearing in the second-order collision
term~\cite{Beneke:2011kc}. For the intensity, they read
\allowdisplaybreaks[4]
\begin{align}
  C^{(2)}_{Ilm}
 & =|\dot{\kappa}| \bigg(- \Delta_{I,lm}^{(2)}(\bm{k}) 
+\delta_{l0}\Delta_{I,00}^{(2)}(\bm{k})  
+4\delta_{l1} v_{\ub,[m]}^{(2)}(\bm{k}) 
+ \delta_{l2}\frac{1}{10}\left[\Delta_{I,2m}^{(2)}(\bm{k}) - 
\sqrt{6}\Delta_{E,2m}^{(2)}(\bm{k}) \right] \notag
\\
&+ \left[A^{(1)}(\bm{k}_1)+\delta_\ub^{(1)}(\bm{k}_1)+
  \delta_{\xe}^{(1)}(\bm{k}_1)\right] \notag \\ & \times 
\bigg\{-\Delta_{I,lm}^{(1)}(\bm{k}_2) 
+\delta_{l0}\Delta_{I,00}^{(1)}(\bm{k}_2)  
+4\delta_{l1} v_{\ub,[m]}^{(1)}(\bm{k}_2)
+  \delta_{l2}\frac{1}{10}\left[\Delta_{I,2m}^{(1)}(\bm{k}_2) - 
\sqrt{6}\Delta_{E,2m}^{(1)}(\bm{k}_2) \right]\bigg\}
\notag \\ & + \sum_{\epsilon = \pm}(-\epsilon) v^{(1)}_{\ub,[m_2]}(\bm{k}_1)
\Delta_{I,(l + \epsilon) m_1}^{(1)}(\bm{k}_2) C^{\epsilon,l}_{m_1 m}
 \notag \\ & + \delta_{l0}\,v^{(1)}_{\ub,[m_2]}(\bm{k}_1)
\left[2\Delta_{I,1m_1}^{(1)}(\bm{k}_2)-4v^{(1)}_{\ub,[m_1]}(\bm{k}_2)\right]
C^{+,0}_{m_1 m} \notag \\
&+ 3\delta_{l1}\,v^{(1)}_{\ub,[m_2]}(\bm{k}_1) \Delta_{I,0m_1}^{(1)}(\bm{k}_2)C^{-,1}_{m_1 m}   
 + \delta_{l2} \,v^{(1)}_{\ub,[m_2]}(\bm{k}_1) \left[7 v^{(1)}_{\ub,[m_1]}(\bm{k}_2)
-\frac{1}{2}\Delta_{I,1m_1}^{(1)}(\bm{k}_2)\right] C^{-,2}_{m_1 m}  
\notag \\
&+\frac{1}{2} \delta_{l3}\,v^{(1)}_{\ub,[m_2]}(\bm{k}_1)
\left[\Delta_{I,2m_1}^{(1)}(\bm{k}_2) -
  \sqrt{6}\Delta_{E,2m_1}^{(1)}(\bm{k}_2) \right]
C^{-,3}_{m_1  m}  \bigg).
\label{eq:2ndorderI}
\end{align}
For the $E$-mode polarisation, they are
\begin{equation}
  \begin{aligned}
C^{(2)}_{Elm} & = |\dot{\kappa}| \bigg(- \Delta_{E,lm}^{(2)}(\bm{k}) -\delta_{l2}
\frac{\sqrt{6}}{10} \left[\Delta_{I,2m}^{(2)}(\bm{k}) -
  \sqrt{6}\Delta_{E,2m}^{(2)}(\bm{k}) \right] 
\\ & +\left[A^{(1)}(\bm{k}_1)+\delta_\ub^{(1)}(\bm{k}_1)+
  \delta_{\xe}^{(1)}(\bm{k}_1)\right] \\ & \times
\bigg\{-\Delta_{E,lm}^{(1)}(\bm{k}_2)  -\delta_{l2} \frac{\sqrt{6}}{10}
  \left[\Delta_{I,2m}^{(1)}(\bm{k}_2) - \sqrt{6}\Delta_{E,2m}^{(1)}(\bm{k}_2)
  \right]\bigg\}
\\
&+\sum_{\epsilon=\pm}(-\epsilon) v^{(1)}_{\ub,[m_2]}(\bm{k}_1)
\Delta_{E,(l + \epsilon)m_1}^{(1)}(\bm{k}_2) D^{\epsilon,l}_{m_1 m} 
\\
&+ \delta_{l2} \frac{\sqrt{6}}{2}v^{(1)}_{\ub,[m_2]}(\bm{k}_1) \left[
  \Delta_{I,1m_1}^{(1)}(\bm{k}_2) -2
  v^{(1)}_{\ub,[m_1]}(\bm{k}_2)\right] C_{m_1 m}^{-,2}
  \\
  & - \delta_{l3}\frac{\sqrt
  6}{2}v^{(1)}_{\ub,[m_2]}(\bm{k}_1)\left[\Delta_{I,2m_1}^{(1)}(\bm{k}_2)
    - \sqrt{6}\Delta_{E,2m_1}^{(1)}(\bm{k}_2) \right] D^{-,3}_{m_1 m} \bigg),
  \end{aligned}
\label{eq:2ndorderE}
\end{equation}
while the $B$-mode terms read
\begin{equation}
  \begin{aligned}
    C^{(2)}_{Blm} & = |\dot{\kappa}| \bigg\{ - \Delta_{B,lm}^{(2)}(\bm{k}) + 
    v^{(1)}_{\ub,[m_2]}(\bm{k}_1)\Delta_{E,lm_1}^{(1)}(\bm{k}_2)
    D^{0,l}_{m_1 m}
    \\
    &- \delta_{l2} \frac{\sqrt{6}}{5}v^{(1)}_{\ub,[m_2]}(\bm{k}_1)
    \left[\Delta_{I,2m_1}^{(1)}(\bm{k}_2) -
      \sqrt{6}\Delta_{E,2m_1}^{(1)}(\bm{k}_2) \right]
    D^{0,2}_{m_1 m} \,\bigg\}.
    \label{eq:2ndorderB}
  \end{aligned}
\end{equation}
In the previous three equations, the quantities $C^{\pm,l}_{m m'}$,
$D^{\pm,l}_{m m'}$ and $D^{0,l}_{m m'}$ are the coupling functions,
which are specific combinations of the Clebsch-Gordan
coefficients. They are given in the appendix A of
Ref.~\cite{Beneke:2011kc}.  Note that we have assumed a cold ensemble
of electrons and are therefore not considering thermal contributions,
such as the thermal Sunyaev-Zel'dovich effect.

In Eqs.~\eqref{eq:2ndorderI}, \eqref{eq:2ndorderE}, and
\eqref{eq:2ndorderB}, the first terms, of the form
$-|\dot{\kappa}|\Delta^{(2)}_{Xlm}$, describe scatterings out of the
line-of-sight and are straightforwardly integrated as an exponential
term in Eq.~\eqref{eq:los}. All the other terms however have to be
accounted for as new collisional sources and add-up to form the term
$\rho_{ab}$ in Eq.~\eqref{eq:los}. In addition, while the sources are
very different for the $E$- and $B$-mode polarisation, these modes do
mix in free-streaming. For practical purposes we therefore do not
refer to $E$- or $B$-mode sources separately, instead they will be
referred to as the polarised sources.

In the following, we discuss the origin of all these quantities and
this allows us to distinguish three categories.

\subsubsection{Pure terms}
\label{sec:pure}
We label as ``pure'' the terms which contain only a single
second-order perturbation. These ones assume a functional form
identical to the linear collision term, $C^{(1)}_{X,lm}$, but with
the second-order perturbations instead of the first order ones. For
the intensity, one obtains from Eq.~\eqref{eq:2ndorderI}
\begin{equation}
C^{(2)}_{I,\pure}
= |\dot{\kappa}| \,\left\{- \Delta_{I,lm}^{(2)}(\bm{k}) 
+\delta_{l0}\Delta_{I,00}^{(2)}(\bm{k})  
+4\delta_{l1} v_{\ub,[m]}^{(2)}(\bm{k}) 
+ \delta_{l2}\frac{1}{10}\left[\Delta_{I,2m}^{(2)}(\bm{k}) - 
\sqrt{6}\Delta_{E,2m}^{(2)}(\bm{k}) \right]\right \} .
\end{equation}
As above-mentioned, the first term accounts for collisions out of our
line-of-sight and has to be discarded from $\rho_{ab}$ as it is
already included in the line-of-sight integration. Next we have a
monopole term\footnote{This term does cancel exactly with the monopole
  part of the first term and precisely appears by enforcing a rigorous
  split between both contributions.} and the generation of a dipole
aligned with the second-order baryon velocity. Finally, there is a
quadrupole emission that also generates $E$-mode polarisation.

Although the pure terms describe the same physics as the well-known
linear collision terms, they do involve the cosmological perturbations
at second order. As such, they can be affected by more complex
phenomena than their linear counterparts. For instance, lensing might
change the multipole moments before reionisation and thereby change
the likelihood of collisions at reionisation and later. As a result,
the pure terms can be affected by all possible second-order sources
prior to the collision and this requires a different numerical
treatment than the other terms discussed below. Let us remark that the
pure terms are defined due to these numerical considerations and do
not represent a specific physical process.

\subsubsection{Gain terms}
\label{sec:gain}
The gain terms describe the generation of perturbations by outgoing
photons that end up being scattered into our line-of-sight. Ignoring
the pure terms in $C^{(2)}_{X}$, the gain terms can be identified by
remarking that they should be quadratic in the linear
perturbations and linked to a given low multipole. The reason being
that photons emitted after a collision follow a simple angular
distribution, dictated by the kinematics of the collision. For Compton
scattering this corresponds to a dipole aligned with the direction of
the electron velocity.

During the evolution of the Universe, the photon power streams to
smaller scales, suppressing the low multipoles, while the baryon
perturbations grow. As a consequence, terms that are quadratic in the
linear baryon perturbations are expected to dominate over the
remaining sources.  As such, the most important gain terms are the
ones that involve the baryon density contrast $\delta_\ub$. These represent the
Sunyaev-Zel'dovich (SZ) effect~\cite{Sunyaev:1980}
\begin{equation}
C^{(2)}_{I,\uSZ}
= |\dot{\kappa}| \delta^{(1)}_\ub(\bm{k}_1) \left\{ 
\delta_{l0}\Delta_{I,00}^{(1)}(\bm{k}) 
+4\delta_{l1} v_{\ub,[m]}^{(1)}(\bm{k})
+ \delta_{l2}\frac{1}{10}\left[\Delta_{I,2m}^{(1)}(\bm{k}) - 
\sqrt{6}\Delta_{E,2m}^{(1)}(\bm{k})\right]\right\}.
\label{eq:CISZ}
\end{equation}
The SZ effect describes the inhomogeneously increased likelihood of
collisions in overdense regions of the Universe. It is particularly
enhanced for very massive, isolated structures, such as the galaxy
clusters~\cite{Yasini:2016pby}. Here we only treat the
  second-order part of this effect, also known as the
  Ostriker-Vishniac effect~\cite{Ostriker:1986, Vishniac:1987, Jaffe:1997ye}, while
  the sources are all the inhomogeneities in the late-time
  Universe. For this reason, we will be referring to it as the diffuse
  SZ effect.

The structure of the diffuse SZ effect is a convolution between the
baryon overdensity and the linear collision gain term. The most
important contribution is the second term involving the baryon
velocity $v_\ub$. This is the so-called kinematic Sunyaev-Zel'dovich
effect (kSZ), which describes an alignment of the outgoing photons
with the baryon velocity. The remaining terms are the generation of a
homogeneous monopole radiation (0SZ) and a quadrupole (2SZ). For the
$E$-mode polarisation we find a similar quadrupole contribution (pSZ),
which describes that an already existing temperature quadrupole may be
re-scattered in the late Universe and converted into polarisation. The
pSZ terms are the diffuse version of the exact same polarization
induced signals that have been discussed for galaxy clusters, see for
instance Refs.~\cite{Sazonov:1999zp, Challinor:1999yz, Cooray:2002cb,
  Bunn:2006mp, Hall:2014wna}.
  
In the standard treatment, the SZ effect is usually first evaluated in
the local rest frame of a galaxy cluster and one has to correct from
its local motion with respect to the CMB frame. For the diffuse SZ
effect, this is not the case as we simulate the entire Universe
already. The above equations have been derived in the frame in which
the CMB dipole has no expectation value. As such, the corrections
describing the impact of the relative motion compared to the local CMB
dipole are very small and would only appear at third-order in
perturbation theory.

As it can be seen from the full Eq.~\eqref{eq:2ndorderI}, the exact same terms
multiplying $\delta_\ub^{(1)}$ in Eq.~\eqref{eq:CISZ} appear convolved with the gravitational potential perturbation
$A^{(1)}$ and the inhomogeneities in the ionised fraction
$\delta_{\xe}^{(1)}$. For the intensity, in addition to SZ, we
therefore have the following gain terms
\begin{equation}
\begin{aligned}
  C^{(2)}_{I,\uEoR} & = |\dot{\kappa}| \delta_{\xe}^{(1)}(\bm{k}_1)  \left\{ 
\delta_{l0}\Delta_{I,00}^{(1)}(\bm{k}) 
+4\delta_{l1} v_{\ub,[m]}^{(1)}(\bm{k})
+ \delta_{l2}\frac{1}{10}\left[\Delta_{I,2m}^{(1)}(\bm{k}) - 
  \sqrt{6}\Delta_{E,2m}^{(1)}(\bm{k})\right]\right\}, \\
C^{(2)}_{I,\uGR} & = |\dot{\kappa}| A^{(1)}(\bm{k}_1)  \left\{ 
\delta_{l0}\Delta_{I,00}^{(1)}(\bm{k}) 
+4\delta_{l1} v_{\ub,[m]}^{(1)}(\bm{k})
+ \delta_{l2}\frac{1}{10}\left[\Delta_{I,2m}^{(1)}(\bm{k}) - 
  \sqrt{6}\Delta_{E,2m}^{(1)}(\bm{k})\right]\right\}.
\end{aligned}
\label{eq:CIEoRandGR}
\end{equation}
The quantity $C^{(2)}_{I,\uEoR}$ comes from the increased likelihood
for collisions in the regions where more free electrons are available
due to patchy reionisation. By analogy with the diffuse SZ gain terms,
we employ the same notation and label the four quantities convolved
with $\delta_{\xe}^{(1)}$ by 0EoR, kEoR, 2EoR, pEoR, respectively.

Although the kinetic and polarisation terms appearing in
$C^{(2)}_{I,\uSZ}+C^{(2)}_{I,\uEoR}$ have been previously
discussed~\cite{Santos:2003, McQuinn:2005ce, Iliev:2006un,
  Dore:2007bz, Dvorkin:2008tf, Park:2013mv, Alvarez:2015xzu,
  Smith:2016lnt}, purely gravitational effects contained in
$C^{(2)}_{I,\uGR}$ are usually omitted. Because they multiply the
lapse perturbation $A^{(1)}$, they describe a mismatch between the
local Minkowski time, corresponding to the physics of the interaction,
and the general relativistic time coordinate governing the evolution
of the photon fluid. Regions of large, or small, $A$ therefore have a
time-accelerated or decelerated interaction rate. By analogy with the
previous expressions, we label the corresponding relativistic effects
as 0GR, kGR, 2GR and pGR.

Finally, there are other gain terms which are not related to an
overall change of the interaction rate. They involve the baryon
velocity multiplying either itself or the photon perturbations at low
multipoles. These terms describe higher order corrections to the
collisional dynamics and encode phase space enhancements for the
outgoing photons beyond the linear structure. Up to our knowledge,
they have been discussed for the first time in Ref.~\cite{Hu:1993tc}
(for the intensity only) and represent a correction to the linear
Collision term, especially to the dominant late time enhanced Doppler
term. Consequently we labelled them DOP in the following. Equivalent
terms also appear in the polarised hierarchies where they generate
polarisation in the outgoing radiation field and we will refer to
these as pDOP.

\subsubsection{Loss terms}
\label{sec:loss}
We define the loss terms as those given by the remaining
contributions which are therefore not uniquely bounded to a given
multipole $l$. They describe collisions out of our line-of-sight and
are thus proportional to the incoming radiation distribution,
involving all multipoles that are present at the moment of
interaction. Their structure is simpler before performing any
multipole decomposition when they are given by
\begin{equation}
C^{(2)}_{ab,\loss} = - |\dot{\kappa}| \left(A^{(1)} + \delta_\ub^{(1)}
  + \delta^{(1)}_{\xe} \right) \Delta^{(1)}_{ab} + |\dot{\kappa}|
\bm{n} \cdot \bm{v}^{(1)}_\ub \Delta^{(1)}_{ab}\,.
\label{eq:Closs}
\end{equation}
The baryon density appears since in overdense regions there are more
potential electrons available for Compton scattering out of the
line-of-sight. For galaxy clusters, this effect is known as the
blurring Sunyaev-Zel'dovich effect~\cite{HernandezMonteagudo:2009ma,
  Yasini:2016pby} and we are dealing with the diffuse version of
it. It will be referred to as bSZ.

The perturbation to the ionisation rate $\delta_{\xe}$ describes the
same effect, but now due to a higher level of ionised fraction in
certain regions of the Universe. It will be refereed to as the blurring
reionisation effect, bEoR.

The term convolved with the lapse function $A$ describes fluctuations
in the relativistic accelerated collision rate, and will be labelled
as the relativistic blurring effect, bGR.

The remaining term is originating from second order phase space
enhancements in the Compton scattering. Collisions out of our
line-of-sight end up being more likely when the velocities of the
baryons are anti-aligned with that direction. We label this novel
contribution as the blurring Doppler effect, bDOP.

All combined, the loss terms account for an effective interaction rate
out of our line-of-sight. As such, all these blurring
effects can only reduce the observed photon intensity but they are
still able to generate polarisation. In particular, when the linear
$E$-mode polarisation is inhomogeneously suppressed, one expects the
resulting photon distribution to acquire a new $B$-mode
contribution. As we show below, the blurring effects significantly
contribute to the overall $B$-mode power spectrum.
 
\section{Numerical implementation}
\label{sec:num}

The previous equations have been solved using the second order
numerical Boltzmann code {\SONG}~\cite{Pettinari:2013he}. Before presenting our
results, in this section, we discuss the numerical challenges that
underlay the calculations of the pure, gain and loss contributions and
how they have been practically overcome.

\subsection{Pure terms}

As discussed in section~\ref{sec:pure}, the pure terms require a
special analysis since they depend on the full second-order
perturbations, which themselves include the response of all prior
non-linear sources. Even though we are dealing with collisions in the
late Universe in this work, these terms necessarily carry information
about various non-collisional sources, such as weak lensing. For this
reason, a direct integration of the pure terms is very
  challenging.

Employing the line-of-sight integration of Eq.~\eqref{eq:los}, all
suppression terms of the form $-|\dot{\kappa}|\Delta_{X,lm}^{(2)}$ in
Eqs.~\eqref{eq:2ndorderI} to \eqref{eq:2ndorderB} can be removed from
the pure source terms and the remaining ones are bound to the low
multipoles $l=0$, $l=1$ and $l=2$ only. This observation is crucial
because it is currently not numerically feasible to solve the
second-order Boltzmann hierarchy up to very large multipole
moments. Instead we employ {\SONG} to solve the full second-order
Boltzmann equations, including all sources, by cutting the hierarchy
at a low multipole, namely $l_{\rm max}=12$. Such a cut is accurate on
small scales as the relevant low multipoles ($l = 0, 1, 2$) are
suppressed by free streaming and could equally be assumed to be
zero. On the large scales, multipoles are only generated slowly and the
lowest multipoles remain accurate over the whole history of the
Universe, independent of the cut value. On the intermediate scales,
where the low multipoles are still relevant but at the same time
dynamically affected by the larger multipoles, we close the hierarchy
at $l_{\rm max}$ to minimise unphysical reflections. Accuracy can be
checked by ensuring the stability of the results with respect to the
multipole at which the cut has been performed.

Once the second-order perturbations have been extracted, induced from both
collisional and gravitational
contributions, we build the line-of-sight sources for the pure terms
and perform the second-order line-of-sight integration with the
{\SONG} code.

\subsection{Gain terms}

As described in section~\ref{sec:gain}, the gain terms are
convolutions over the linear perturbations and are the diffuse version
of the SZ, EoR, GR and DOP effects. All the involved quantities are
bounded to the lowest multipoles and, as a consequence, are in all
points similar to sources that are already included in the {\SONG}
code for the study of recombination\footnote{At recombination, all
  larger multipoles are suppressed due to the tight coupling between
  electrons and photons.}. First-order perturbations in {\SONG} are
computed from the linear solver {\CLASS} and we have employed already
existing routines to build the line-of-sight sources and integrate
them to the present time. The numerical properties of the gain terms
are however very different from the recombination sources included in
{\SONG} and some code optimisations have been required to actually
perform the computations around the EoR.

\subsection{Loss terms}

The loss terms are not generating new perturbations but modify the
existing linear ones\footnote{This includes the conversion of $E$- to
  $B$-mode polarisation.}. The line-of-sight sources presented in
section~\ref{sec:loss} involve, a priori, all multipole moments at the
EoR or later. Including the loss terms directly in {\SONG}, as we have
done for the gain terms, ends up being numerically not feasible due to
the large multipole moments present in the late Universe. However, one
can notice that the blurring effects act in a very similar way to
lensing and this allows us to derive an analytical expression for
them.

\subsubsection{Blurring potential}
\label{sec:kappablur}

Before multipole decomposition, the differential equation for the loss
terms reads
\begin{equation}
\dot{\Delta}^{(2)}_{\loss,ab} + i \bm{n} \cdot \bm{k}
\Delta^{(2)}_{\loss,ab} = -|\dot{\kappa}|\Delta^{(2)}_{\loss,ab} +
\rho^{(2)}_{\loss,ab} \,,
\end{equation}
where $\rho^{(2)}_{\loss}$ encompasses all the collisional sources for
the loss terms that run over all multipole moments. From
Eq.~\eqref{eq:Closs}, one gets
\begin{equation}
\rho^{(2)}_{\loss,ab} = - |\dot{\kappa}| \left(A^{(1)} +
\delta_\ub^{(1)} + \delta^{(1)}_{\xe} -\bm{n} \cdot \bm{v}^{(1)}_\ub
\right) \Delta^{(1)}_{ab} \,.
\end{equation}
Employing the line-of-sight integration~\eqref{eq:los}, we find
\begin{equation}
  \begin{aligned}
& \Delta^{(2)}_{\loss,ab}(\eta,k) = \int_{\etaini}^{\eta} \dd \eta' e^{-
  i \bm{n} \cdot \bm{k} (\eta - \eta') - \kappa(\eta,\eta')}
C^{(2)}_{\loss,ab}(\eta',\bm{k}_1,\bm{k}_2) \\  & =
\int_{\etaini}^{\eta} \dd \eta' e^{- i \bm{n} \cdot \bm{k} (\eta -
  \eta')-\kappa(\eta,\eta')} \left|\dot{\kappa}(\eta')\right| \left[\bm{n} \cdot
  \bm{v}_\ub^{(1)}(\bm{k}_1) - A^{(1)}(\bm{k}_1) -
  \delta^{(1)}_\ub(\bm{k}_1) - \delta^{(1)}_{\xe}(\bm{k}_1) \right] \Delta^{(1)}_{ab}(\bm{k}_2) \,.
  \end{aligned}
\end{equation}
Next we replace the distribution function
$\Delta^{(1)}_{ab}(\bm{k}_2)$ by its own linear line-of-sight
integration. In order to obtain a tractable final expression, one can
make the approximation of considering only the recombination sources
for the linear perturbations. In other words, the impact of the
late-ISW effect is relocated to recombination. Notice that the same
approximation is employed in the lensing analysis and found to
slightly distort only the largest angular scales. Doing so, one gets
\begin{equation}
\Delta^{(1)}_{ab}(\eta,k,\bm{n}) \simeq \int_{\etaini}^{\eta} \dd \eta' e^{- i \bm{n} \cdot \bm{k} (\eta - \eta')-\kappa(\eta,\eta')} \rho^{(1)}_{\rec,ab}(\eta',\bm{k})\,.
\end{equation}
Making the convolution explicit and disentangling both line-of-sight
integrations yields
\begin{equation}
  \begin{aligned}
 \Delta^{(2)}_{\loss,ab}(\eta,k) &=
 \int_{\etaini}^{\eta} \negthinspace \dd \eta'
\int_{\etaini}^{\eta'} \negthinspace \dd \eta'' \int  \negthinspace
    \dfrac{\dd^3\bm{k}_1}{(2\pi)^3}\dfrac{\dd^3\bm{k}_2}{(2\pi)^3}
    \left|\dot{\kappa}(\eta')\right|  e^{- i \bm{n} \cdot \bm{k} (\eta - \eta')-\kappa(\eta,\eta')}  e^{- i \bm{n} \cdot \bm{k}_2 (\eta' - \eta'')-\kappa(\eta',\eta'')}  \\ 
& \times  (2\pi)^3 \delta^{3}\left(\bm{k}-\bm{k}_1-\bm{k}_2 \right)
    \bigg[\bm{n} \cdot \bm{v}^{(1)}_\ub(\eta',\bm{k}_1) -
      A^{(1)}(\eta',\bm{k}_1) - \delta^{(1)}_\ub(\eta',\bm{k}_1) \\ & -
      \delta^{(1)}_{\xe}(\eta',\bm{k}_1) \bigg] \rho^{(1)}_{\rm{rec},ab}(\eta'',\bm{k}_2) \\ 
&= \int_{\etaini}^{\eta} \negthinspace \dd \eta'
    \int_{\etaini}^{\eta'} \negthinspace \dd \eta''
    \int \negthinspace \dfrac{\dd^3\bm{k}_1}{(2\pi)^3}
    \dfrac{\dd^3\bm{k}_2}{(2\pi)^3} \left|\dot{\kappa}(\eta')\right| e^{- i \bm{n} \cdot \bm{k}_1 (\eta - \eta')}  e^{- i \bm{n} \cdot \bm{k}_2 (\eta - \eta'') -\kappa(\eta,\eta'')}  \\
& \times  (2\pi)^3 \delta^{3}(\bm{k}-\bm{k}_1-\bm{k}_2)
    \bigg[\bm{n} \cdot \bm{v}^{(1)}_\ub(\eta',\bm{k}_1) -
      A^{(1)}(\eta',\bm{k}_1) - \delta^{(1)}_\ub(\eta',\bm{k}_1) \\ & -
      \delta^{(1)}_{\xe}(\eta',\bm{k}_1) \bigg] \rho^{(1)}_{\rec,ab}(\eta'',\bm{k}_2)\,.
\end{aligned}
\end{equation}
This equation can be separated into one integration over
$\bm{k}_1$, taking into account the perturbations to the collision
rate $\bm{n} \cdot \bm{v}^{(1)}_\ub - A^{(1)} - \delta^{(1)}_\ub -
\delta^{(1)}_{\xe}$ and a second integration over $\bm{k}_2$ involving
the usual linear collision sources $\rho^{(1)}_{\rec,ab}$. A further
simplification is to extend the integration domain of $\eta''$ to the
final time $\eta$ since the linear sources are only present around
recombination and are assumed to vanish afterwards. One may now
replace the second line-of-sight integration with $\Delta_{ab}$, running up
to the final time $\eta$ instead of stopping at $\eta'$ (or
reionisation), i.e.
\begin{equation}
\begin{aligned}
  \Delta^{(2)}_{\loss,ab}(\eta,k)
&=  \int \dfrac{\dd^3\bm{k}_1}{(2\pi)^3} \dfrac{\dd^3\bm{k}_2}{(2\pi)^3}
  \int_{\etaini}^{\eta} \dd \eta'  e^{- i \bm{n} \cdot \bm{k}_1 (\eta
    - \eta')} (2\pi)^3 \delta^{3}(\bm{k}-\bm{k}_1-\bm{k}_2)
  \left|\dot{\kappa}(\eta') \right| \\ & \times \left[\bm{n} \cdot
    \bm{v}^{(1)}_\ub(\eta',\bm{k}_1) - A^{(1)}(\eta',\bm{k}_1) -
    \delta^{(1)}_\ub(\eta',\bm{k}_1) -
    \delta^{(1)}_{\xe}(\eta',\bm{k}_1) \right] \Delta^{(1)}_{ab}(\eta,\bm{k}_2)\,.
\end{aligned}
\end{equation}
This naturally leads to the definition of a \emph{blurring potential} (by
analogy with the lensing potential), dependant on the blurring sources
\begin{equation}
\begin{aligned}
  \kappa^{(1)}_{\blur}(\eta,\bm{k}_1) & \equiv -\int_{\etaini}^{\eta} \dd \eta' e^{- i
  \bm{n} \cdot \bm{k}_1 (\eta - \eta')} \left|\dot{\kappa}(\eta')
\right| \\ & \times \left[\bm{n} \cdot \bm{v}^{(1)}_\ub(\eta',\bm{k}_1) - A^{(1)}(\eta',\bm{k}_1) -
  \delta^{(1)}_\ub(\eta',\bm{k}_1) -
\delta^{(1)}_{\xe}(\eta',\bm{k}_1) \right],
\end{aligned}
\end{equation}
from which we find the simple relation 
\begin{equation}
\Delta^{(2)}_{\loss,ab}(\eta,k) =  -\int \dfrac {\dd^3
  \bm{k}_1}{(2\pi)^3} \dfrac{\dd^3 \bm{k}_2}{(2\pi)^3} (2\pi)^3
\delta^{3}(\bm{k} -\bm{k}_1-\bm{k}_2) \kappa^{(1)}_{\rm blur}(\eta,\bm{k}_1) \Delta^{(1)}_{ab}(\eta,\bm{k}_2)\,,
\end{equation}
or in real space 
\begin{equation}
\Delta^{(2)}_{\loss,ab} =  -\kappa^{(1)}_{\blur} \Delta^{(1)}_{ab} .
\label{eq:2ndblur}
\end{equation}
The blurring potential $\kappa^{(1)}_{\blur}$ allows us to keep track
of the likelihood of collisions in the late Universe out of our
line-of-sight and we can then apply it directly on the linear present
day photon distribution function. The blurring potential can be seen
as the inhomogeneous extension of the homogeneous background optical
depth $\kappa$ and may be included in the line-of-sight integration in
the exact same way, i.e.
\begin{equation}
\Delta_{\loss,ab} =  \exp\left(-\kappa^{(1)}_{\blur} \right) \Delta^{(1)}_{ab} \approx \Delta^{(1)}_{ab} -  \kappa^{(1)}_{\blur} \Delta^{(1)}_{ab}.
\end{equation}
The first term reproduces the linear result while the second term is
the correction due to the non-linear collisional loss terms. Our
relation encompasses the expression employed
in Ref.~\cite{Dvorkin:2008tf} for the study of the bEoR effect. Here
it also contains the bSZ, bGR and bDOP contributions.

As noted before, we have neglected the impact of the late-ISW
effect on the collisional loss term at reionisation. This may
distort the largest multipoles but it should be noted that
the very same approximation is used in the usual lensing
treatment~\cite{Lewis:2006fu, Fidler:2014zwa}. In principle, since the
late-ISW affects only the largest moments, we could have included it
directly in the {\SONG} code together with the gain terms. We have however not done so
as comparable contributions in the gain terms turn out to be
suppressed.

\subsubsection{Structural similarity with weak lensing}

In real space, Eq.~\eqref{eq:2ndblur} is structurally similar to the
one involved in weak lensing, where a lensing potential deflects
photons. The blurring potential does not change the photon direction,
but reduces the intensity of the photons (scattering out of the
line-of-sight). Hence the derivatives present in the lensing framework
are absent here. Apart from this difference, we may employ almost the
same weak lensing equations for the blurring~\cite{Bartelmann:1999yn,Lewis:2006fu,
Fidler:2014zwa}.

More precisely, for weak lensing, the leading order expression reads
\begin{equation}
\Delta^{(2)}_{\lens} = \nabla^a \Psi^{(1)}_{\lens} \nabla_a \Delta^{(1)},
\end{equation}
while for the blurring we have a similar relation but without the angular
derivatives $\nabla_a$, i.e.,
\begin{equation}
\Delta^{(2)}_{\loss} = -\kappa^{(1)}_{\blur} \Delta^{(1)} .
\end{equation}
In the flat sky limit, which is accurate enough on scales
  $l>10$, one gets for the intensity
\begin{equation}
\Delta_{\loss,I}^{(2)}(l) =- \int \frac {\dd^2 l'}{2\pi}
\kappa^{(1)}_{\blur}(l-l') \Delta^{(1)}_{I}(l')\,.
\end{equation}
Assuming that the correlations between $\kappa_{\blur}$ and $\Delta$
are suppressed, we obtain the angular power spectra 
\begin{equation}
\mathcal{C}_{\loss,I}(l) = \int \frac {\dd^2 l'}{2\pi} \mathcal{C}^{\kappa}(l-l') \mathcal{C}_{I}(l').
\end{equation}
For polarisation, in the flat sky limit, we may also employ the
lensing results after minimal modifications to obtain
\begin{equation}
\begin{aligned}
\mathcal{C}_{\loss,E}(l) &= \int \dfrac {\dd^2 l'}{2\pi} \mathcal{C}^{\kappa}(l-l') \mathcal{C}_E(l') \cos^2 \left[2 \left(\phi_{l'} - \phi_{l}\right)\right],\\
\mathcal{C}_{\loss,B}(l) &= \int \dfrac {\dd^2 l'}{2\pi} \mathcal{C}^{\kappa}(l-l') \mathcal{C}_B(l') \sin^2\left[2\left(\phi_{l'} - \phi_{l}\right)\right],
\end{aligned}
\end{equation}
where $\phi_l$ denotes angular coordinate on the flat sky and governs
the conversion of $E$- into $B$-modes.

In lensing, the intensity $I$ and $E$-mode polarisation receive a
correction from a first-order times third-order {\cross{1}{3}}
term~\cite{Bartelmann:1999yn,Lewis:2006fu}. By expanding the
exponential of the blurring potential up to second order, we obtain an
analogous term for the blurring effect. It represents multiple
(correlated) interactions during reionisation and we find
\begin{equation}
  \begin{aligned}
    \mathcal{C}_{\loss,I}(l) &= \int \dfrac{\dd^2 l'}{2\pi} \mathcal{C}^{\kappa}(l-l') \mathcal{C}_{I}(l') - \mathcal{C}_{I}(l) \int \dfrac {\dd^2 l'}{2\pi} \mathcal{C}^{\kappa}(l')\,, \\
\mathcal{C}_{\loss,E}(l) &=  \int \dfrac{\dd^2 l'}{2\pi}
\mathcal{C}^{\kappa}(l-l') \mathcal{C}_E(l') \cos^2\left[2
  \left(\phi_{l'} - \phi_{l}\right)\right] - \mathcal{C}_{E}(l) \int
\dfrac{\dd^2 l'}{2\pi} \mathcal{C}^{\kappa}(l') \,,\\
\mathcal{C}_{\loss,B}(l) &= \int \dfrac{\dd^2 l'}{2\pi} \mathcal{C}^{\kappa}(l-l') \mathcal{C}_B(l') \sin^2\left[2\left(\phi_{l'} - \phi_{l}\right)\right].
  \end{aligned}
\label{eq:blur}
\end{equation}
In addition to the power spectra we may also derive the induced
bispectrum
\begin{equation}
\left \langle \Delta_I(l_1)\Delta_I(l_2)\Delta_I(l_3) \right
\rangle_{\loss} = \dfrac{1}{2\pi} \delta^2(l_1 + l_2 +
l_3)\left(C_{l_1}^{I \kappa} C^I_{l_2}  \, l_1\cdot l_2 +
\circlearrowleft \right) ,
\end{equation}
where $\circlearrowleft$ denotes permutations of $l_1$, $l_2$ and
$l_3$. It should be noted that, as for lensing, the bispectrum is suppressed from the weak correlation between the late time blurring potential and the early Universe linear photon perturbations. 

In conclusion, the blurring potential can be computed in the same way
as the lensing potential in a typical linear Boltzmann code. The only
complication comes from the $\bm{n} \cdot \bm{v}_\ub$ term which
enters as a dipole source, as opposed to the monopole sources for the
lensing potential. We have implemented the modifications required to
compute the blurring potential $\kappa_\blur$ in the linear Boltzmann
code {\CLASS}.

\subsection{Epoch of Reionisation}

Although most of the new sources can be computed directly from the
{\SONG} and {\CLASS} codes, the only exception is the perturbation of
the ionisation fraction $\delta_{\xe}$. The physics of reionisation is
intrinsically non-linear and beyond the scope of these
codes~\cite{Barkana:2006ep}. The details of the epoch of reionisation
have been explored using numerical simulations, see for instance
Refs.~\cite{McQuinn:2007dy, Zahn:2006sg, Jelic:2008jg, Zahn:2010yw,
  Iliev:2015aia, Lin:2015bcw, Bauer:2015tta}.  These techniques are
out of the scope of the present work although we still need order of
magnitude accurate predictions for the power spectra involving
perturbations in the ionised fraction. For this reason, we have
employed the same method as in Refs.~\cite{Mao:2008ug, Clesse:2012th},
namely the use of analytic fitting formulae to the numerical
simulations of Refs.~\cite{McQuinn:2007dy, Zahn:2010yw}. In
particular, for the power spectrum of the ionised fraction, we have
used the following expression
\begin{equation}
  \xet^2 P_{\delta_{\xe}}(\eta,k) = N(\xet) \left(1 - \xet \right)^2 \left\{1 +
\alpha(\xet) k R(\xet) + \left[k R(\xet) \right]^2 \right\}^{-\gamma(\xet)/2} P_{\delta_{\ub}}(\eta,k),
\end{equation}
which is parametrised by the background normalised ionised fraction
$\xet(\eta)\equiv \xe(\eta)/\xe(\eta_0)$, $\xe(\eta_0)$ being the
total ionised fraction once the Universe is completely
reionised\footnote{In the {\CAMB} and {\CLASS} codes, this quantity
  can be greater than one due to Helium reionisation, i.e. $\xe$ is
  defined as the hydrogen ionised fraction.}. The length scale
$R(\xe)$ typically keeps track of the ionised bubble size while
$P_{\delta_{\ub}}(\eta,k)$ is the baryon power spectrum whose
evolution is solved employing {\CLASS}. Notice that $\xe(\eta)$ is
also completely determined by the homogeneous reionisation model
implemented in the {\CLASS} and {\CAMB} codes~\cite{Lewis:2008wr}. The
functional form for the amplitude $N(\xe)$, bubble sizes $R(\xe)$, the
shape parameters $\alpha(\xe)$ and $\gamma(\xe)$, have been chosen as
minimal power law expansions in $\xe$ to be regular enough and to
match the values reported in Ref~\cite{Mao:2008ug}. Let us stress that
only $P_{\delta_{\xe}}(\eta,k)$ is required in
Eqs.~\eqref{eq:2ndorderI} to \eqref{eq:2ndorderB} and we do not need
to specify the cross-spectrum $P_{\delta_{\xe} \delta_{\ub}}(\eta,k)$.
\begin{figure}
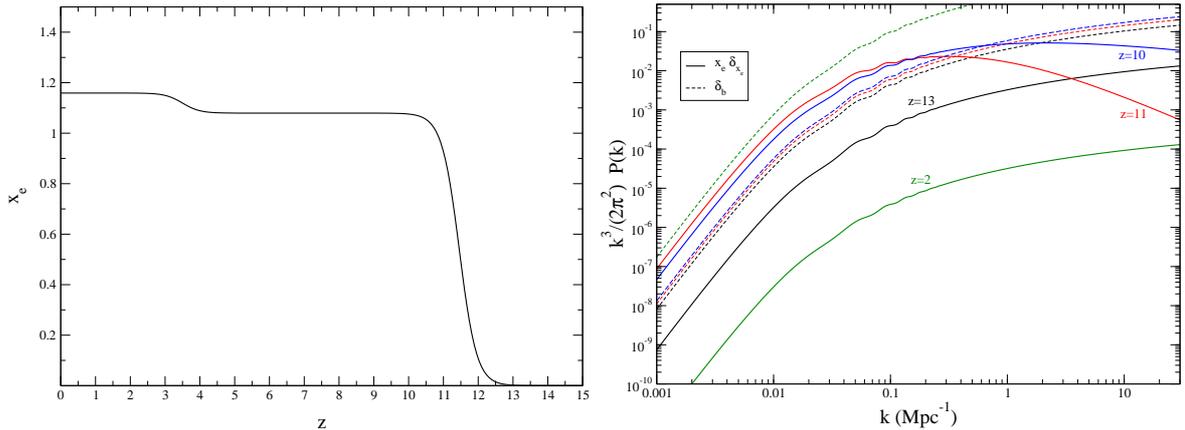

  \begin{center}
    \includegraphics[width=\twofigw]{xetot}
    \includegraphics[width=\twofigw]{pxexereio}
    \caption{Left panel: total background ionised fraction as a function of
      redshift used. The right panel shows the power spectra for $\xe
      \delta_{\xe}$ as modelled in our analysis at different redshifts
    (thick solid). For comparison, we have also represented the
      power spectra for the baryon overdensity (dashed thin curves),
      which steadily grows in the late Universe.}
    \label{fig:reioxe}
  \end{center}  
\end{figure}
As an illustration, we have represented in figure~\ref{fig:reioxe} the
evolution of $\xe(\eta)$, $\xe^2 P_{\delta_{\xe}}(\eta,k)$ and
$P_{\delta_{\ub}}(\eta,k)$ at different times during the EoR for the
currently favoured {\LCDM} model by the Planck satellite
collaboration~\cite{Ade:2013zuv, Paoletti:2016vtg}.

\section{Results}
\label{sec:res}

As above-mentioned, all the {\LCDM} cosmological parameters have been
fixed to their currently favoured values by the Planck satellite
collaboration~\cite{Ade:2013zuv, Paoletti:2016vtg}, and we have only
considered primordial scalar perturbations. We have used the {\SONG}
code to compute the late Universe collisional sources.

At second-order
in perturbation theory we may compute the sources in isolation when
studying individual perturbations $\Delta^{(2)}_{Xlm}$. However,
when computing power spectra, correlations between the different
effects may become relevant. This is the case for the various parts of
the collision terms, and more generally, also between collision
sources and other secondaries. For this reason, we have fully included
the correlations between the gain and pure terms and, separately, the
correlations within the loss terms. However, correlations with
non-collisional secondaries, and between the gain and loss term, have
been neglected so far. 

Let us notice that, in principle, the power spectra are given by a
genuine second-order perturbation {\cross{2}{2}} and the correlation
of a linear perturbation with a third order one {\cross{1}{3}}. Third
order perturbations are beyond the scope of the {\SONG} code and
cannot be straightforwardly computed. However, we expect the
{\cross{2}{2}} part to dominate the power spectra for the following
reasons.  For some effects, such as the kSZ or DOP for example, the
second-order perturbations introduce new structures that are enhanced
compared to the linear order. If the related third-order terms are not
further enhanced compared to the already large second order ones, we
find that $\cross{2}{2} \gg \cross{1}{3}$. See also the related
discussion based on geometrical considerations in
Ref.~\cite{Hu:1993tc}. On the other hand, for the blurring effects,
the potentially dominant third-order terms would be multiple blurring
events along the line-of-sight, generated by the same structures that
already exist at second order. We have been able to include these
higher order contributions by performing multiple applications of the
blurring potential, which is equivalent to what is performed in the
more usual lensing computation (see section.~\ref{sec:kappablur}).

Finally, let us stress that for the most interesting $B$-mode
polarisation calculations, the linear part vanishes and the
{\cross{1}{3}} part is thus completely absent. Our derivation of the
$B$-mode power spectrum is therefore exact.

\subsection{Loss terms}
In the literature the EoR loss term has been discussed in details,
while we provide an unified approach further including the bSZ, bGR and
bDOP effects.  We first summarise our results for the blurring
potential, followed by the calculation of the induced non-linear power
spectra.

\subsubsection{Blurring potential}

\begin{figure}
\begin{center}
  \includegraphics[width=\onefigw]{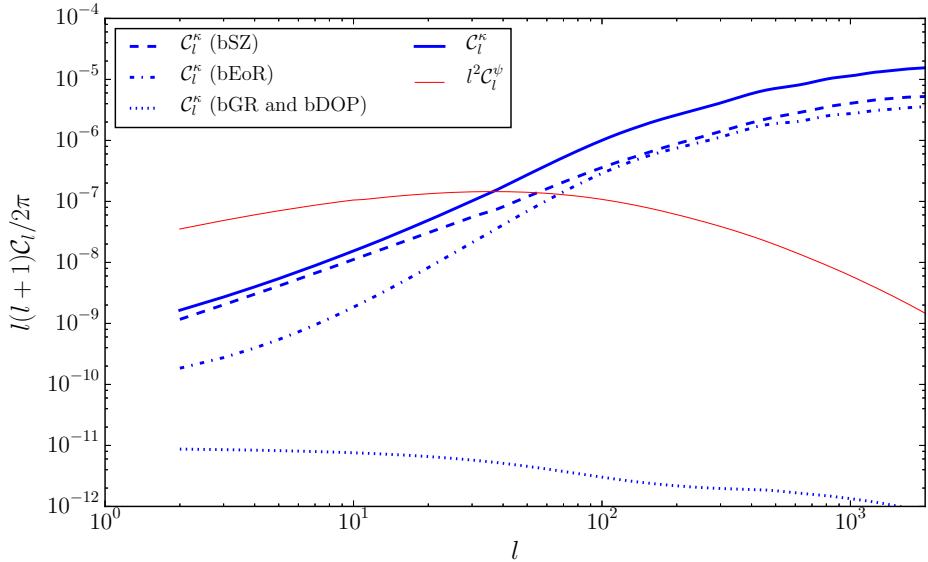}
  \caption{Angular power spectrum $\mathcal{C}_l^{\kappa}$ of the
    blurring potential $\kappa_{\blur}$ in blue together with its main
    components. For comparison, we have represented the lensing power
    spectrum $\mathcal{C}_l^{\Psi}$, multiplied by $l^2$. It should be
    noted that the lensing does act on the derivatives of the linear
    perturbations while the optical depth acts on the perturbations
    themselves and an absolute comparison cannot be drawn from this
    plot.}
    \label{fig:Repot}
\end{center}
\end{figure}

The loss terms are accounted for by the blurring potential, which has
been represented in figure~\ref{fig:Repot}, separated into its diverse
components. As can be checked on this figure, the relativistic (bGR)
and second order phase space enhancements (bDOP) effects have a
negligible impact on $\kappa_{\blur}$, which is entirely dominated by
the blurring Sunyaev-Zel'dovich and blurring reionisation effects, bSZ
and bEoR, respectively.

This is expected for the relativistic blurring, bGR, sourced by the
lapse function $A$ that is much smaller than the late Universe baryon
densities responsible for the bSZ effect. The bDOP depends on the
orientation of the velocity and tends to cancel along the
line-of-sight, at least on the larger scales studied in this
work. Concerning the bEoR contribution, as can be seen in
figure~\ref{fig:reioxe}, $\delta_{\xe}$ can become larger than the
baryon overdensity at EoR, but vanishes later on. After integrating
along the line-of-sight, we find that bSZ and bEoR are comparable in
amplitude. Only on the largest length scales the contribution from
reionisation appears to be suppressed with respect to bSZ and this can
be explained by the power streaming to smaller multipoles after the
EoR.

As previously mentioned, the bEoR contribution has been studied in
some details in Ref.~\cite{Dvorkin:2008tf}. Our result shows that it
is a good approximation to not include the bGR and bDOP
contributions. In addition, we find that the bSZ and bEoR effects are
about $50\%$ correlated. The reason is that the physics of
reionisation is directly linked to the baryon over-densities, which
are also responsible for the bSZ effect. A separate analysis of the
bSZ and bEoR contributions would find the total blurring power
spectrum significantly smaller.

The angular power spectrum of the blurring potential is almost
featureless (blue solid curve in figure~\ref{fig:Repot}). All
existing structures in the baryon overdensities, such as the baryon
acoustic oscillations, are smoothed out by the integration along the
line-of-sight.

Whereas the lensing potential is peaked at scales around $l \simeq 10$
to $100$, we find the blurring potential to have support up to much
smaller scales. Lensing does locally distort the linear CMB
fluctuations according to the large scale gravitational potentials,
while blurring represents a reduction of power from regions where
collisions out of the line-of-sight are more likely, tracing the
distribution of matter in the Universe. This imprints a small scale
signature onto the already inhomogeneous incoming CMB leading to a
rise of power on the very small scales. As a consequence, the blurring
must eventually dominate over lensing in the damping tail where the
primary CMB does not posses much power.

\subsubsection{Blurring induced CMB anisotropies}
\label{sec:blurTB}

\begin{figure}
  \begin{center}
    \includegraphics[width=\onefigw]{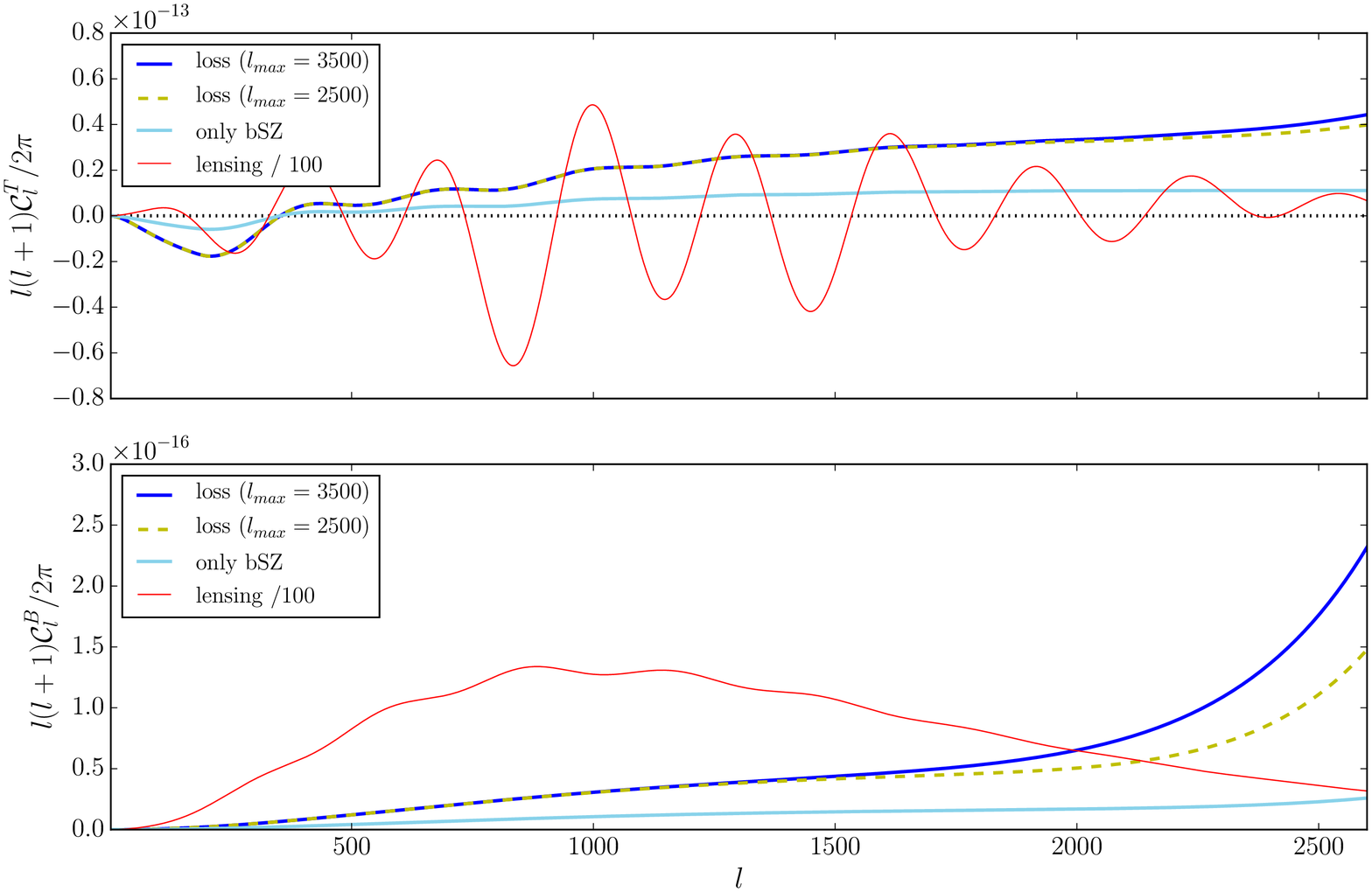}
    \caption{The blurring temperature ($\calC_l^T)$ and $B$-mode
      ($\calC_l^B$) angular power spectra. The light blue line shows
      the bSZ effect in isolation. It has to be compared
      to the entire blurring spectrum (dark blue curve). Correlations
      cannot be neglected, in particular between bSZ and bEoR. We have
      also represented the lensing power spectra (divided by a factor
      $100$) in red. The blurring effects appear to be relatively
      enhanced on the smaller scales while the lensing becomes
      suppressed. This plot also shows the impact of the multipole
      moment cuts, either at $l=3500$ or $l=2500$.}
    \label{fig:blurTB}
  \end{center}
\end{figure}

As discussed in section~\ref{sec:loss}, we have integrated
equations~\eqref{eq:blur} to compute the temperature and $B$-mode
angular power spectra induced by the blurring effects. They have
been represented in figure~\ref{fig:blurTB}.

Let us notice that all signatures of the linear spectra are washed out
and the spectra extends far into the Silk damping tail. As previously
mentioned, this is due to the support of $\kappa_{\blur}$ on the small
scales.  The light blue curve in figure~\ref{fig:blurTB} represents
the angular spectra obtained by considering only the bSZ contribution,
while the dark blue line includes all blurring contributions. As we
have already seen in the blurring potential, the bSZ and bEoR effects
are of comparable magnitude and strongly correlated.

For the temperature power spectra, the blurring effects are comparable
in amplitude to one percent of the lensing up to $l\simeq 2000$. Let
us notice that the lensing and blurring effects are possibly strongly
correlated as they are sourced by the same density fields. The only
difference is that the lensing depends, via the potential, on the
inverse Laplacian of the density. Such a correlation might affect the
lensing up to the $10\%$ level, especially towards the smaller
scales and could potentially be responsible of some of the Planck
lensing anomalies~\cite{Ade:2015zua}.
As previously discussed, we have kept the collision term in
isolation and cannot currently compute this effect. We leave a
combined lensing and blurring analysis for a future work.

Since we only perform the analysis to second order, we
test the dependence of our results on the small scales by cutting the
integration of Eq.~\eqref{eq:blur} at $l=2500$ instead of $l=3500$. The temperature power
spectrum is relatively stable under this cut, which suggests that our
analysis should not be affected by unaccounted small scale non-linear
physics beyond second order.

In addition to modifying the intensity, the blurring terms also
generate polarisation. By locally reducing the linear $E$-modes due to
collisions out of the line-of-sight, the blurring induces a mix of
$E$- and $B$-modes. The $B$-mode angular power spectrum has been
represented in the lower panel of figure~\ref{fig:blurTB}. As for the
temperature, the blurring $B$-modes spectrum is almost featureless and
comparable to a percent of the lensing generated
$B$-modes. Interestingly, $B$-modes are more significantly affected by
non-linear corrections, as visible deviations appear by changing the
multipole cut around $l=1500$. This is caused by the lack of large
scale power in the linear $E$-mode polarisation, enhancing the
signal's dependence on the smaller scales. Again, unaccounted
correlations between lensing and blurring $B$-modes might contaminate
the $B$-modes after lensing has been cleaned and we let their
derivation for a future work.

\subsection{Pure and Gain terms}

The pure and gain terms can be evaluated with the {\SONG} code (see
sections~\ref{sec:pure} and \ref{sec:gain}). Up to our knowledge, the
computation of the pure terms has not been performed so far and we
find that they remain always negligible compared to the much larger
gain terms, typically by at least two orders of magnitude. As long as
the perturbative approach is applicable, the second-order
perturbations remain smaller than their linear counterparts, at least
when the later are not vanishing. This also holds for the growing
baryon perturbations. The pure terms have the same structure as the
linear collision ones and are therefore bound to be perturbatively
small, provided we remain focused on the large and mildly non-linear
scales.

For the first time we provide a full analysis of all second order gain
sources.  While many of the gain terms are comparable in amplitude to
the pure terms, a few are significantly larger and dominate the
signal. For the intensity these are the kSZ and the kEoR
effects. Polarisation is not directly induced by the kSZ effect and
various other contributions then become relevant. We find the
polarisation gain terms to be driven by the pSZ, pDOP and pEoR
effects. As one could have intuitively guessed, the dominant sources
end up being the ones that contain the maximum number of baryon
perturbations.

\subsubsection{Accuracy tests}

\begin{figure}
  \begin{center}
    \includegraphics[width=\onefigw]{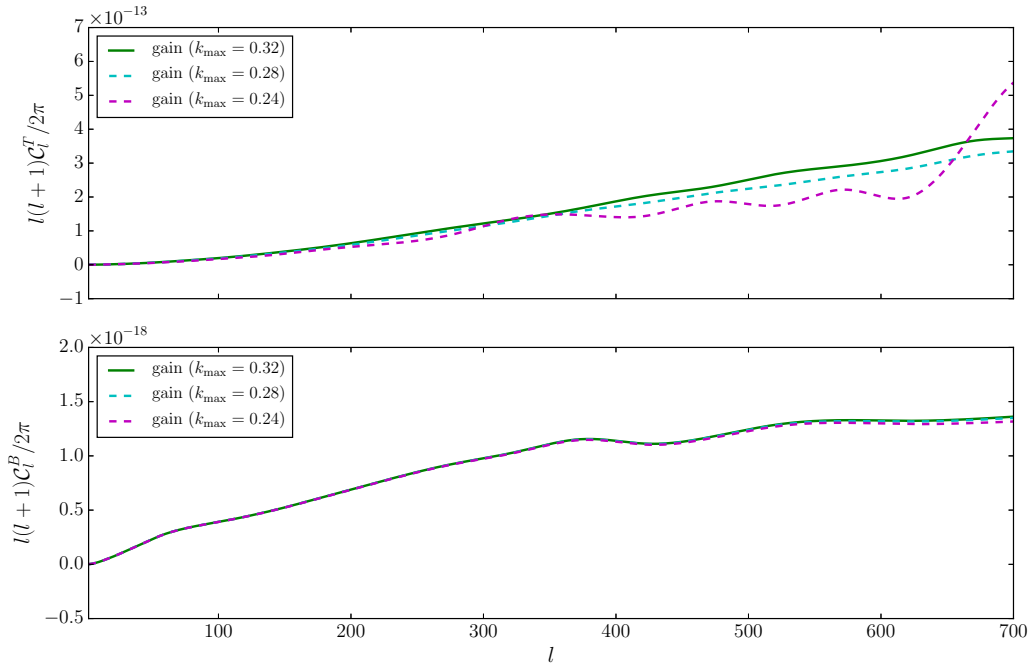}
    \caption{Accuracy tests for the gain terms angular power spectra
      associated with a truncation of the Fourier decomposition at
      different wavenumber values $k_{\max}$ (in $\Mpc^{-1}$). Since
      {\SONG} is a second-order perturbation code, the fully
      non-linear corrections expected at small scales are not
      accounted for and one can trust the numerical results only when
      these ones decouple from the small scale modes. We find that
      multipoles larger than $l=500$ already receive significant power
      from the small scale modes and are beyond the capabilities of a
      second-order perturbative analysis. It should be noted that we
      have optimised our numerical parameters for the range $l < 500$
      and the output cannot be trusted on smaller scales.}
    \label{fig:kmax}
  \end{center}
\end{figure}

In Fig.~\ref{fig:kmax}, we analyse how much power is transferred from
the very small and non-linear scales into our large scale angular
power spectra by cutting the Fourier integrations at a given
wavenumber $k_{\max}$.

For the temperature (upper panel), rather small multipoles, of the
order $l \simeq 500$, appear to be already affected by small scale
physics. This comes from the kSZ effect, which is given by a
convolution over the baryon density contrast and velocity
perturbations. Both are present over the entire range of scales but
gravitational instability make them of larger amplitude at the smaller
scales. When evaluating the convolution product at large scales, the
small scale modes do contribute significantly and physically describe
the transfer of power from two small scale modes into a large scale
perturbation. We conclude that a second-order code is not suitable to
compute kSZ and kEoR effects for multipoles larger than $l \gtrsim
500$, where non-linear corrections become relevant. Our result
can therefore be trusted for $l < 500$.

As can be seen in the $B$-mode power spectrum in
figure~\ref{fig:kmax}, for polarisation, the situation is improved and
non-linearities are not expected to play a significant role up to
$l=700$. Contributing to polarisation, the pSZ and pEoR effects,
depend on the linear photon quadrupoles, and consequently one Fourier
mode is evaluated at very large scales. Even though the baryon density
is then evaluated at a smaller scale, the transfer of power from small
scales into large scales is suppressed. However, polarisation is also
generated by the phase space enhancement terms (pDOP), which involve
the baryon velocities squared and allow a transfer of power similar to
the kSZ terms.

\subsubsection{Gain induced CMB anisotropies}

\begin{figure}
\begin{center}
  \includegraphics[width=\onefigw]{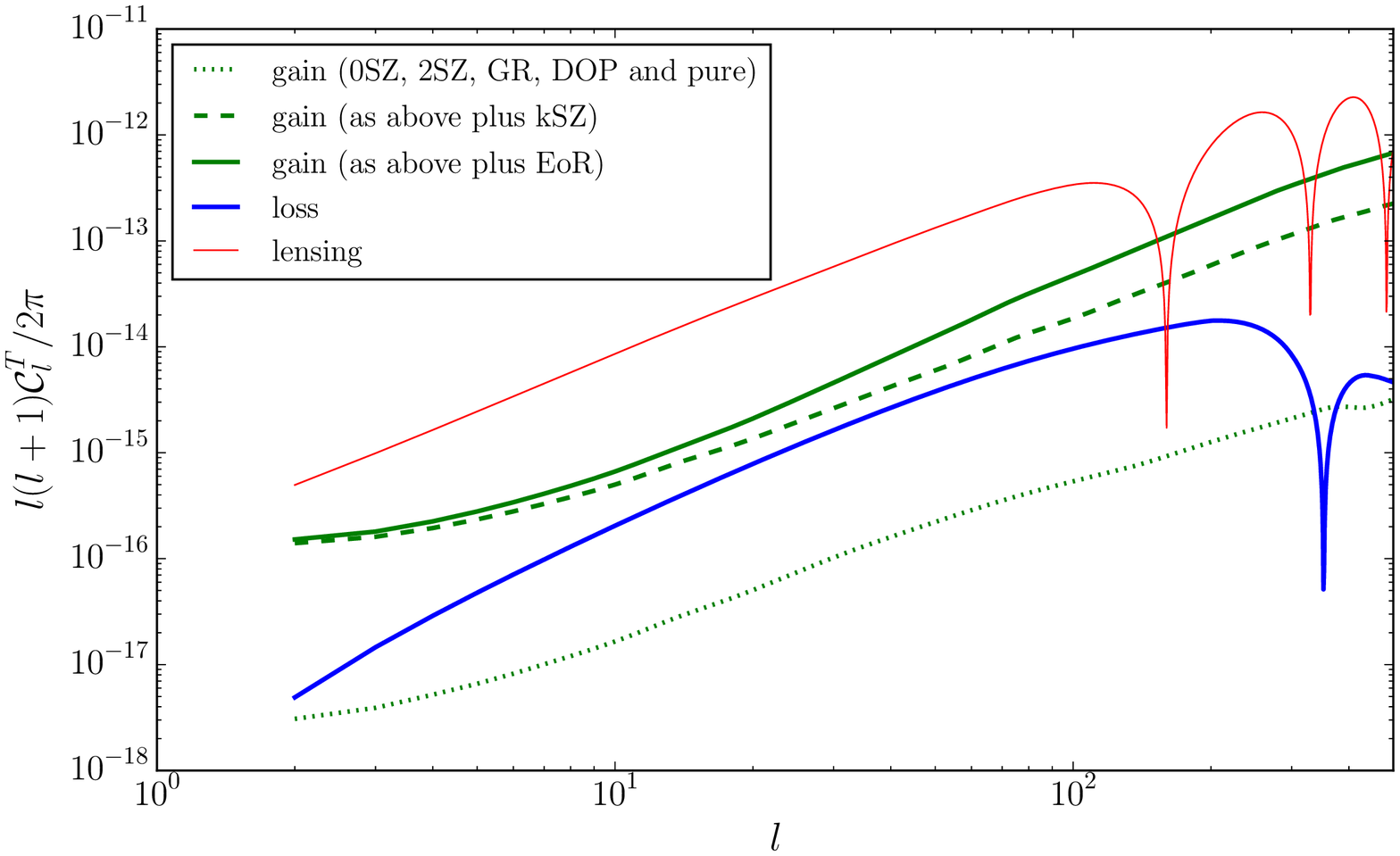}
  \includegraphics[width=\onefigw]{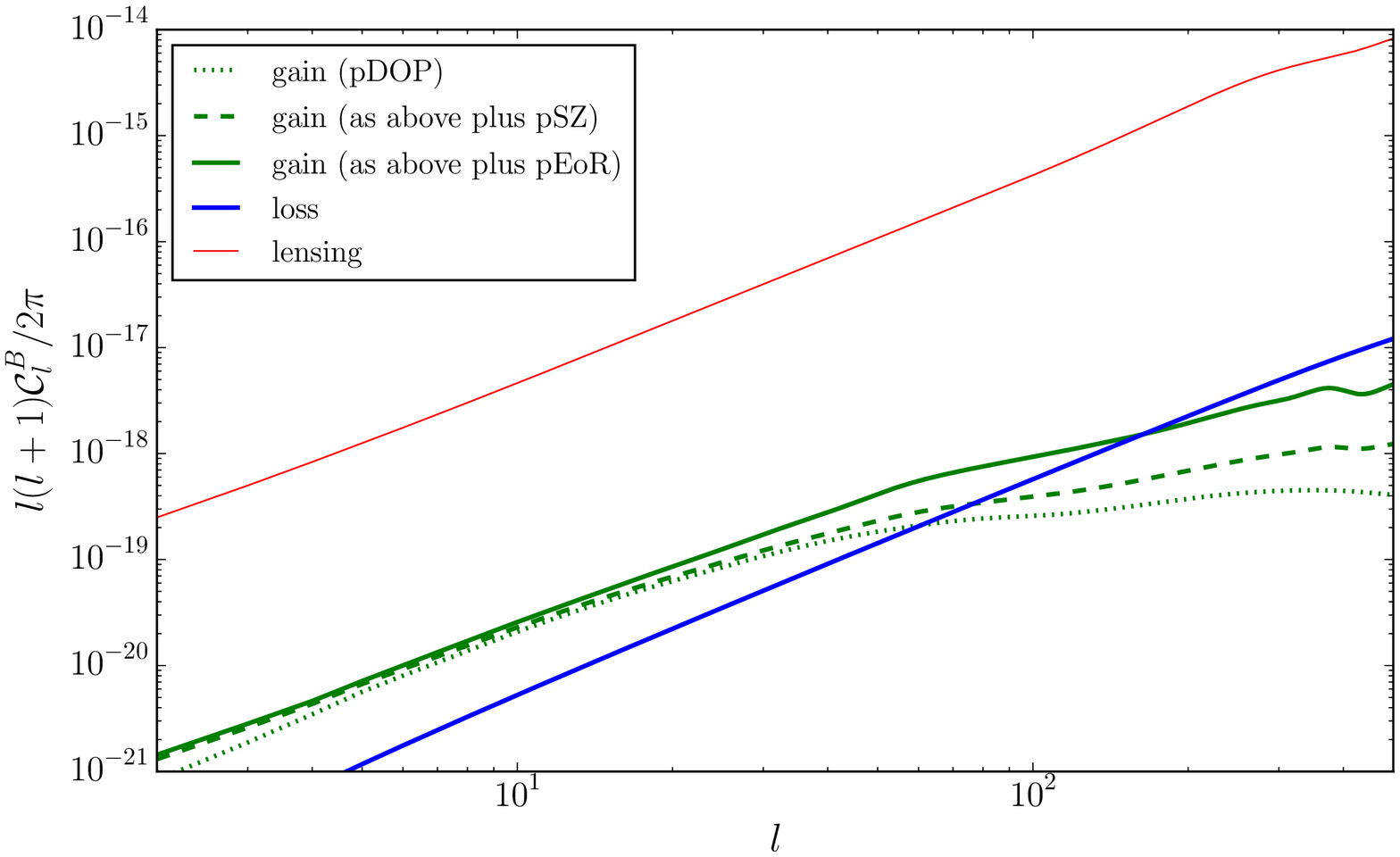}
  \caption{Temperature (upper panel) and $B$-mode (lower panel)
    angular power spectrum for the gain terms, separated into their
    main components. For comparison, the blurring (loss terms) and
    lensing have been reported. For temperature, the signal is
    dominated by the kSZ and kEoR effects, while the relativistic
    (GR), pure and phase space enhancement (DOP) contributions remain
    two orders of magnitude smaller. For polarisation, the large
    scales are dominated by the pDOP effect, while on smaller scales
    the pSZ and pEoR effects become relevant. The signal is comparable
    in size to the loss contributions. }
   \label{fig:gainTB}
\end{center}
\end{figure}

The temperature angular power spectrum $l(l+1)\calC_l^T/(2\pi)$
generated by the gain and pure terms has been represented in the upper panel of
figure~\ref{fig:gainTB}. We find that the contributions due to the
relativistic (GR), phase space enhancement (DOP) and pure terms are
subleading, and the signal is essentially made up of the kSZ and kEoR
effects~\cite{McQuinn:2005ce, Iliev:2006un, Alvarez:2015xzu,
  Park:2013mv}. The kEoR is smaller than the kSZ on very
  large scales as it tends to be erased during the post EoR
  period. As for the blurring, we find a strong correlation between
kSZ and kEoR increasing their combined power spectrum. Overall, the
gain contribution for the temperature power spectrum is about one
order of magnitude larger than the blurring. It remains smaller than
the lensing over the entire range of analysed scales, but not by
much. It is possible that correlations
between the gain and loss terms, or the gain and lensing terms are important and
these have not been considered here.

The lower panel of figure~\ref{fig:gainTB} shows the induced $B$-mode
angular power spectrum $l(l+1)\calC_l^B/(2\pi)$ induced by the gain
terms. In addition to the pSZ and pEoR effects, which have been
studied in isolation in Refs.~\cite{Sazonov:1999zp, Santos:2003, Dore:2007bz}, we
have been able to compute for the first time the pGR, pDOP and pure
contributions together with their correlations. As opposed to the
temperature power spectra, which are dominated by the kSZ and kEoR
effects, we find the pDOP contributions to be important for the
$B$-modes, where they actually dominate the signal on the large scales. This
can be understood by remarking that, contrary to the pSZ effect, the
pDOP terms can transfer power from the small scales to the large
scales. In parallel, the pSZ and pEoR effects become more important on
the smaller scales and are, again, strongly correlated.

\begin{figure}
\begin{center}
    \includegraphics[width=\onefigw]{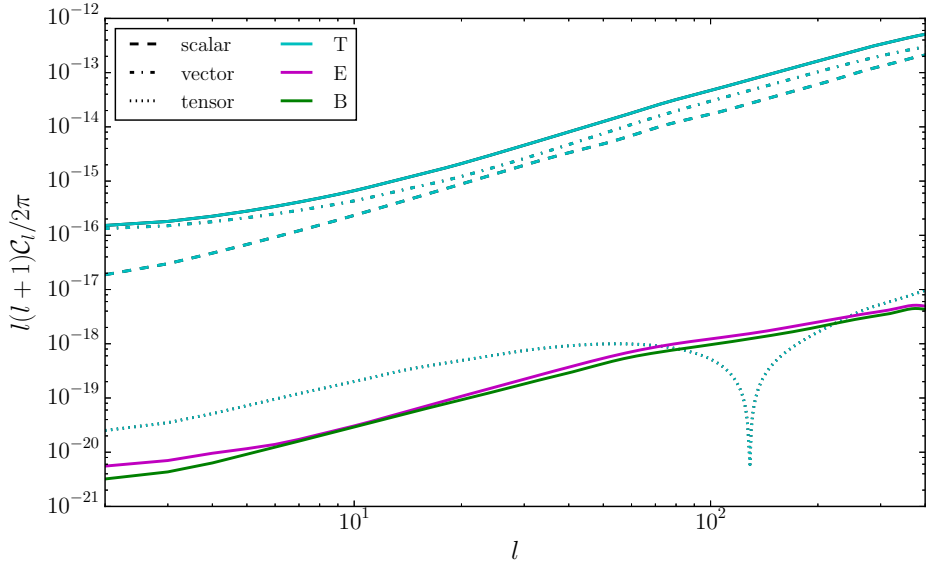}
    \caption{Temperature and polarisation spectra induced from the
      collisional gain terms. The scalar, vector and tensor
      contributions of $\calC_l^T$ are represented. Temperature
      dominates over polarisation due to the kSZ and kEoR effects,
      which are able to generate scalar and vector modes only. The
      much smaller tensor modes are however comparable to both $E$ and
      $B$ polarisation which mix in free streaming.}
    \label{fig:gainSVT}
\end{center}
\end{figure}

Figure~\ref{fig:gainSVT} finally shows the temperature, $E$- and
$B$-mode angular power spectra induced by the gain terms and we have
separated $\calC_l^T$ into its scalar ($m = 0$), vector ($m = \pm 1$)
and tensor ($m = \pm 2$) contributions. The dominant kSZ and kEoR
effects driving $\calC_l^T$ only induce scalars and vectors, while the
tensors are sourced by the subleading 2SZ and DOP
terms. Polarisation is also not induced from the kSZ and kEoR effects,
which explains why $\calC_l^E$ and $\calC_l^B$ are of similar
magnitude as the temperature's tensors. The $E$- and $B$-modes mix in
free-streaming and consequently have a similar shape and
amplitude. This is reason why we have not represented the $E$-mode
signal in the previous plots. Only on very large scales, one can see
that $\calC_l^B$ is slightly lower than $\calC_l^E$. On these scales,
there is indeed not enough time after EoR to convert all of the
induced $E$-mode polarisation into $B$-modes.

\subsection{Combined result of collisional secondaries}

\begin{figure}
\begin{center}
  \includegraphics[width=\onefigw]{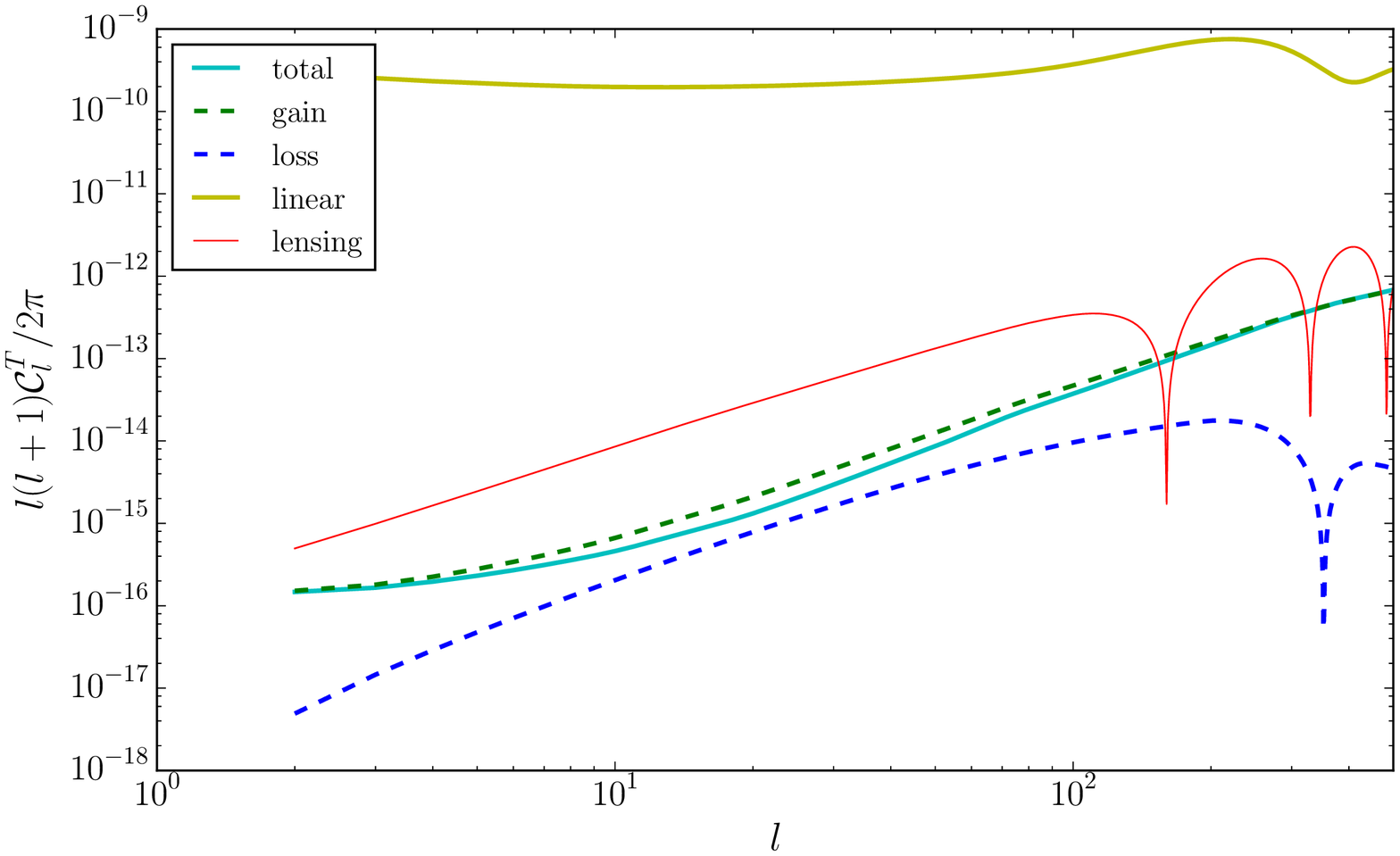}
  \includegraphics[width=\onefigw]{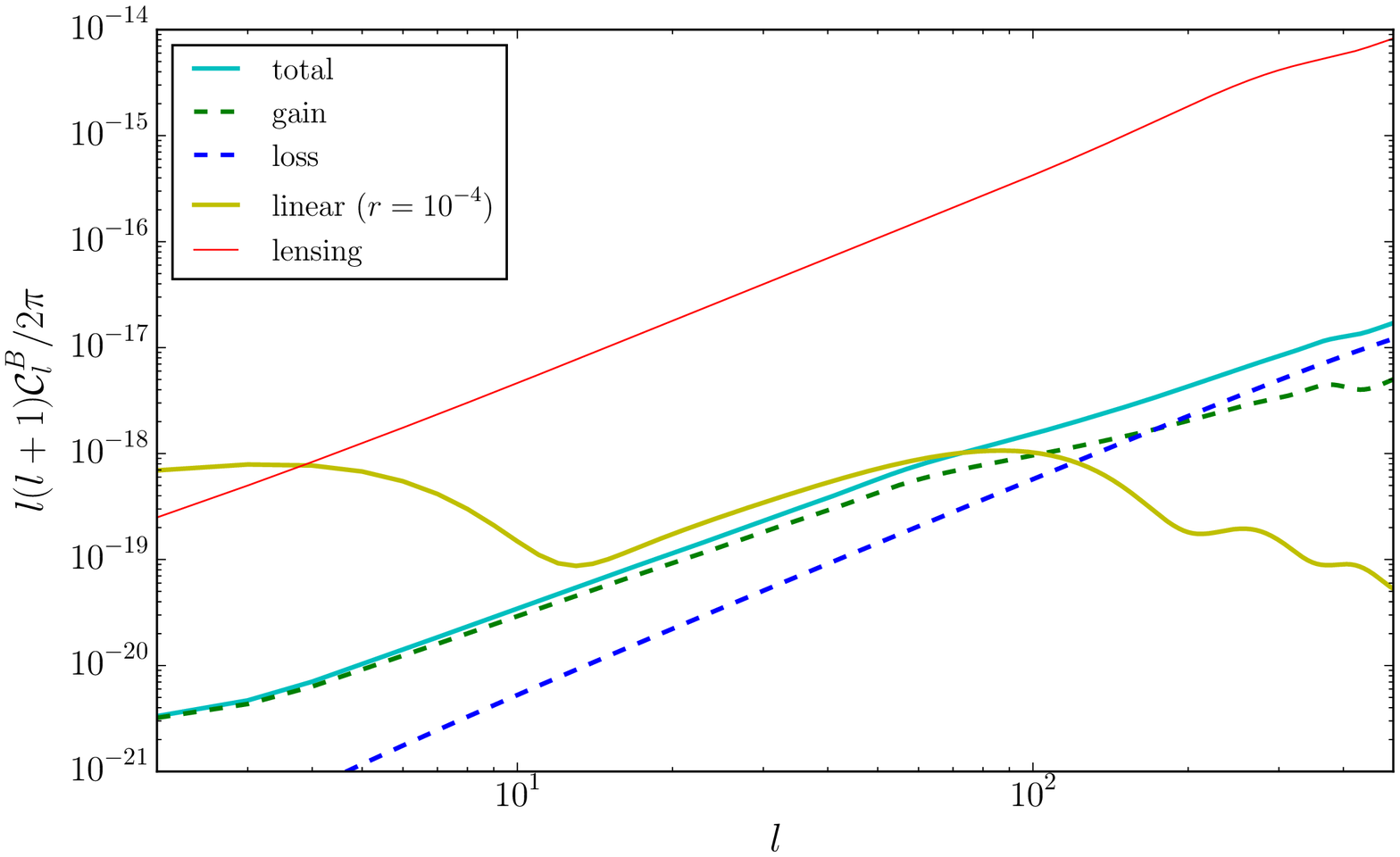}
  \caption{Temperature (upper panel) and $B$-mode (lower panel)
    angular power spectra induced by all terms. For comparison, we
    have represented the lensing power spectra as well as primordial
    $B$-modes that would be generated by a tensor-to-scalar ratio
    $r=10^{-4}$. For temperature, $C_l^T$ remains smaller than lensing
    but only by a factor of a few. For polarisation, the induced
    $B$-modes remain always smaller than lensing but match the
    primordial ones all over the range $l \in [10,100]$.}
  \label{fig:totalTEB}
\end{center}
\end{figure}

In figure~\ref{fig:totalTEB}, we have summarized our results and
plotted the total temperature and $B$-mode polarisation angular power
spectra for all collisional secondaries. For comparison, the lensing
induced signals and the primordial $B$-modes associated with a
tensor-to-scalar ratio of $r=10^{-4}$ have been represented. The
temperature power spectrum $\calC_l^T$ is essentially driven by the
gain contributions on all scales and remains smaller than lensing, but
only by a factor of a few. The polarisation spectra $\calC_l^E$ and
$\calC_l^B$ are always of comparable amplitude due to their mixing by
free streaming and we have represented only $\calC_l^B$. They are
sourced by the gain terms on large scales and by the blurring on small
scales. On all the multipole range studied, the polarisation power
spectra remain much smaller than the lensing induced ones.

Let us stress however that the $B$-mode signal we have computed is of
comparable amplitude to the primordial one induced by a
tensor-to-scalar ratio of $r = \order{10^{-4}}$, all over the range $l
= 10$ to $100$. Various experiments have been proposed to target very
small tensor-to-scalar ratios~\cite{Matsumura:2013aja,
  Abazajian:2016yjj}, and they would rely on our ability to perform
delensing on the foreground-cleaned $B$-modes~\cite{Smith:2010gu,
  Errard:2015cxa, Carron:2017vfg, Millea:2017fyd}. Our result shows
that, from $l = 10$ to $100$, all of the late Universe \emph{diffuse}
secondaries are important, and especially the pDOP
contribution. Interestingly, it remains a clean window for primordial
$B$-modes on the very large scales, $l<10$, favouring the case for
full sky experiments~\cite{Ishino:2016wxl, 2017arXiv170604516D, Abitbol:2017nao}.

Finally, as already mentioned, the $B$-modes presented in our plots
are not subject to {\cross{1}{3}} terms and the approximation of
including all possible contributions up to second order is very
accurate on the large scales; fully non-linear corrections eventually
appearing for $l \gtrsim 800$. Nevertheless, some uncertainties remain
from the physics of reionisation and the currently neglected
correlations between gain, loss and lensing, which will be
investigated in a future work.
 
\section{Conclusions}\label{sec:con}

We have employed the second-order Boltzmann code {\SONG} to obtain an
unified framework to compute the CMB anisotropies induced by late-time
collisional secondaries. These were the last missing pieces of a
complete second order Boltzmann treatment of the cosmological
perturbations~\cite{Pettinari:2014iha}. Our results show that the
collisional sources in the late Universe are significantly larger than
the second order recombination effects already implemented in {\SONG}
and studied in Refs.~\cite{Beneke:2011kc, Pettinari:2014iha}. While
the likelihood for a CMB photon to have an interaction in the late
Universe is much smaller than the chance of having last scattered at
recombination, the imprint in the CMB is nevertheless enhanced. This
is due to the baryon perturbations, which grow during the evolution of
the Universe and are much larger at reionisation compared to
recombination.

For the first time, we have included all collisional secondaries in a
relativistic non-linear Boltzmann code. Our sources include the
diffuse versions of the Sunyaev-Zel'dovich effect for the temperature
and polarisation, the impact of patchy reionisation, but also the
blurring Sunyaev-Zel'dovich. Our framework naturally encompasses a
range of novel relativistic and pure second order effects, plus some
phase space enhancement terms which have only been discussed for
temperature in Ref.~\cite{Hu:1993tc}.

Among these novel contributions, the pure terms are structurally very
complex as they include non-linear effects from the evolution of the
Universe before the EoR. However, we have found that they are of the
expected size for second-order perturbative effects and thus remain
subleading compared to the others. The same holds for the relativistic
corrections, labelled as bGR, 0GR, 2GR and pGR, that depend on the
gravitational potentials. We further show that the polarised phase
space enhancement terms, referred to as pDOP, are relevant and even
dominate the large scale B-mode polarisation power spectrum.

The kinetic and blurring Sunyaev-Zel'dovich and reionisation effects
describe the inhomogeneously increased likelihood of collisions in
overdense regions and due to patchy reionisation.  While the kSZ, kEoR
and bEoR effects are well-known in the literature (see
section~\ref{sec:intro}), we also compute the polarised blurring and
find that it provides a significant contribution to the induced
$B$-mode polarisation.
 
Reionisation and SZ effects are often discussed in isolation. Having
both effects implemented in the {\SONG} code allows us to study their
correlations. The SZ and EoR contributions have been shown to be
strongly correlated, by no less than $50\%$, which significantly affects
the resulting power spectra compared to a separate analysis.

Due to the similarities between lensing and blurring that we have
pointed out, one may expect a comparable correlation between the
lensing and blurring effects. We have found that the blurring is
typically comparable to a percent of the lensing signal but
correlations could possibly be as large as $10\%$ of the lensing
signal. These ones have however not been computed and will be pursued
in a future work.

In summary, we have shown that the $B$-modes generated at, and after,
EoR, including the novel effects discussed above, are of the same
amplitude than the primordial ones stemming from a tensor-to-scalar
ratio of $r=10^{-4}$ over a large range of scales. They may become
relevant in future CMB experiments. Potential correlations of these
with the lensing $B$-modes might further enhance the signal.

Concerning the numerics, we have tested the impact of the non-resolved
small scale perturbations in {\SONG} on the larger scale power spectra
and find that our accuracy is maintained up to $l \simeq 500$ for the
gain sources, while the blurring is accurate up to $l \simeq 1500$.
The ability to compute all late Universe collisional secondaries will
be soon added to the publicly available
\href{https://github.com/coccoinomane/song}{{\SONG}}
code~\cite{Pettinari:2013he}. In the meanwhile the code is available
upon request.

\acknowledgments
This work is supported in part by the Belgian Federal Science Policy
Office through the Inter-university Attraction Pole P7/37.

\bibliographystyle{JHEP}
\bibliography{references}

\end{document}